\documentclass[preprint,showpacs,preprintnumbers,amsmath,amssymb]{revtex4}

\usepackage{graphicx}
\def\bm #1{\bbox{#1}}

\begin{document}


\title{Theoretical investigation of moir\'e patterns in quantum images}

\author{M. P. Almeida, P. H. Souto Ribeiro\\Instituto de F\'\i sica - Universidade Federal do Rio de Janeiro\\
Caixa Postal 68528, Rio de Janeiro - RJ, 21941-972, Brazil.}


\author{J. A. O. Huguenin, A. Z. Khoury\\
Instituto de F\'\i sica - Universidade Federal Fluminense\\
Niter\'{o}i - RJ, 24210-340, Brazil.}


\date{\today}

\begin{abstract}

Moir\'e patterns are produced when two periodic structures with different spatial
frequencies are superposed. The transmission of the resulting
structure gives rise to spatial beatings which are called moir\'e fringes.
In classical optics, the interest in moir\'e fringes comes from the fact that
the spatial beating given by the frequency difference gives information about
details(high spatial frequency) of a given spatial structure.
We show that moir\'e fringes can also arise in the spatial distribution of the
coincidence count rate of twin photons from the parametric down-conversion,
when spatial structures with different frequencies are placed in the path of
each one of the twin beams.
In other words,we demonstrate how moir\'e fringes can arise from quantum images.

\end{abstract}

\maketitle

Quantum image is the name used by several authors of scientific reports,
for the spatial correlations of a  light source
possessing some non-classical property. Even though in some cases the calculated
or measured pattern has a classical counterpart, it is known that
some {\em quantum images} are actually spatial patterns with no classical
counterpart. The activity in this field has increased over the last ten years\cite{qimages}.

Besides the academic interest in demonstrating differences between classical
and quantum descriptions of light, it would be of great importance to demonstrate
the possibility of some practical application. One way towards practical
applications is the combination between quantum images and other known physical
processes. One example is the use of quantum spatial correlations for improving
the resolution of imaging systems. Apodization in correlated images \cite {apod} and the so called quantum lithography, whose fundamental
principle was demonstrated by Fonseca et al.\cite{sebast} and has
received this name afterwards\cite{shih1,boyd1,dowling}. Though promising, none
of these practical applications has overcome all the technical difficulties in actually
replacing classical systems with some advantage. In the case of the quantum
lithography, for instance, no substrate has been found to efficiently absorb
two photons.

In this report, we would like to introduce the theoretical analysis for other promising practical application for quantum images. It has been
demonstrated experimentally\cite{jaugusto} that it is possible to observe
moir\'e patterns in quantum images, for two fundamentally different experimental configurations. One configuration
is based on the transfer of the angular spectrum from the pump to the twin
photons\cite{monken} and other is based directly on the non-local correlations
between signal and idler photons\cite{nldslit}. In both cases, the calculations demonstrate
that the coincidence count rate depends on a product of functions describing
the periodic structures used. A good
agreement between the theory presented here and the experimental results 
of Ref.\cite{jaugusto} is achieved

\section{Pump-Idler Setup}
In this section we will calculate the coincidence profiles for the two configurations mentioned above. We start from the state of the field generated by spontaneous parametric down-conversion for a thin crystals, in the monochromatic and paraxial approximation \cite{monken}

\begin{equation}
\left | \psi \right\rangle\,=\, c_1\left | vac \right\rangle + c_2 \int d {\bm q}_s \int d {\bm q}_i {\mathcal V_{z_0}}\left({\bm q}_s+{\bm q}_i\right) \left | 1,{\bm q}_s \right\rangle  \left | 1,{\bm q}_i \right\rangle,
\label{eq1}
\end{equation}
where $c_1$ and $c_2$ are constant coefficients, $\left | 1,{\bm q}_s \right\rangle$ , $\left | 1,{\bm q}_i \right\rangle$ are the single photon Fock states with transverse wave vectors corresponding to the down-converted signal and idler modes, respectively. $\left | vac \right\rangle$ represents the vacuum state of the electromagnetic  field. $ {\mathcal V_{z_0}}\left({\bm q}_s+{\bm q}_i\right)$, with ${\bm q}_p\,=\, {\bm q}_s+{\bm q}_i$, is the angular spectrum of the pump, at the crystal position. In terms of the angular spectrum immediately after the object ${\mathcal{V}_0}({\bm q}_p)$, we can write:
\begin{equation}
\label{eq2}
{\mathcal V}_{z_0}\propto {\mathcal V}_{0}\exp\left[-i\frac{q_p^2z_1}{2k_p}\right].
\end{equation}
The coincidence count rate $C\left({\bm \rho}_s, {\bm \rho}_i \right)$ , in the transverse plane of the detectors, is proportional to the fourth-order correlation function $G^{(2,2)}\left({\bm  \rho}_s, {\bm \rho}_i \right)$, given by:
\begin{eqnarray}
\label{eq3}
G^{(2,2)}\left({\bm  \rho}_s, {\bm \rho}_i \right)=
\left\langle \psi \right| \hat{E}^{(-)}({\bm \rho}_s) \hat{E}^{(-)}({\bm \rho}_i)\hat{E}^{(+)}({\bm \rho}_s) \hat{E}^{(+)}({\bm \rho}_i) \left|\psi \right\rangle=&&\nonumber\\
\left|\left\langle vac\right|\hat{E}^{(+)}({\bm \rho}_s) \hat{E}^{(+)}({\bm \rho}_i) \left|\psi \right\rangle\right|^2.
\end{eqnarray}
\begin{figure}[t]
\centering
\includegraphics*[width=8cm]{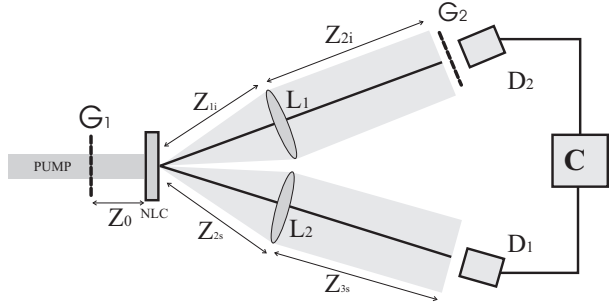}
\vspace{0.25cm}
\caption{Pump-Idler setup.}
\label{f1}
\end{figure}
First we will consider the setup showed in the Fig. \ref{f1}. The first grating is placed
in the pump beam and its image is transferred to the transverse coincidence
distribution between the signal and idler beams, with the aid of lenses L1 and
L2\cite{malmeida}. The second grating is placed in the idler beam. The resulting moir\'e pattern is produced by the superposition of the conditional image and the second grating. We call this configuration pump-idler setup. Assuming that $ {\mathcal A}_2$ is the transmission function of the grating $G_2$, we can write the field operator for the idler mode as:
\begin{eqnarray}
\label{eq4}
\hat{E}_{i}^{(+)}\left({\bm \rho}_i\right)={\mathcal A}_2\left({\bm \rho}_i\right)\int d{\bm \rho}''_i\,\int d{\bm \rho}'_i \int d{\bm q}_i'\hat{a}\left({\bm q}_i'\right) \exp \left[i{\bm q}_i\cdot {\bm \rho'}_i \right]&&\nonumber\\
\times \exp\left[i\left(\left |{\bm \rho''}_i-{\bm \rho'}_i\right |^2\,\frac{k_i}{2z_1i}\right)\right]{\mathcal T}\left({\bm \rho''}_i\right) \exp\left[i\left(\left |{\bm \rho}_i-{\bm \rho''}_i\right |^2\,\frac{k_i}{2z_2i}\right)\right],&&\nonumber\\
\end{eqnarray}
where  $k_i$ is the wave number of the idler beam.
${\mathcal T}_1\left({\bm \rho}_i''\right)$
is the transmission function of the lens $L_1$, given by:
\begin{equation}
{\mathcal T}_1\left({\bm \rho}_i''\right)=\exp\left(-i\frac{\rho''^2_i\,k_i}{2f}\right).
\label{eq5}
\end{equation}
The field operator for the signal mode, at the detection plane can be written as:
\begin{eqnarray}
\label{eq6}
\hat{E}_{s}^{(+)}\left({\bm \rho}_s\right)&=&\int d{\bm \rho}''_s\,\int d{\bm \rho}'_s \int d{\bm q}_s'\hat{a}\left({\bm q}_s'\right)\exp\left[i{\bm q'}_s\cdot{\bm \rho'}_s\right] \nonumber\\
&&\times \exp\left[i\left(\left |{\bm \rho''}_s-{\bm \rho'}_s\right |^2\,\frac{k_s}{2z_1s}\right)\right]{\mathcal T}\left({\bm \rho''}_s\right)\\
&& \times \exp\left[i\left(\left |{\bm \rho}_s-{\bm \rho''}_s\right |^2\,\frac{k_s}{2z_2s}\right)\right]. {}
\nonumber
\end{eqnarray}
\par
Using Eqs.(\ref{eq1})-(\ref{eq6}) we can calculate the fourth order correlation function:
\begin{eqnarray}
\label{eq7}
\left\langle vac\right|\hat{E}^{(+)}({\bm \rho}_s) \hat{E}^{(+)}({\bm \rho}_i) \left|\psi \right\rangle &=&\nonumber\\
{\mathcal A}_2\left({\bm \rho}_i\right)\int d{\bm \rho''}_s\,\int d{\bm \rho'}_s \int d{\bm \rho''}_i\,\int d{\bm \rho'}_i
\int d{\bm q'}_s\int d{\bm q'}_i{\mathcal V}_{z_0}\left({\bm q'}_s+{\bm q'}_i\right)&&\nonumber\\
\times \exp\left[i\left(\left |{\bm \rho''}_s-{\bm \rho'}_s\right |^2\,\frac{k_s}{2z_{1s}}\right)\right]\exp\left(-i\frac{\rho''^2_s\,k_s}{2f}\right)&& \nonumber\\
\times \exp\left[i\left(\left |{\bm \rho}_s-{\bm \rho''}_s\right |^2\,\frac{k_s}{2z_{2s}}\right)\right]\exp\left(i{\bm q'}_s{\bm \rho'}_s\right)&&\nonumber\\
\times \exp\left[i\left(\left |{\bm \rho''}_i-{\bm \rho'}_i\right |^2\,\frac{k_i}{2z_1}\right)\right]\exp\left(-i\frac{\rho''^2_i\,k_i}{2f}\right)&&\nonumber\\
\times \exp\left[i\left(\left |{\bm \rho}_s-{\bm \rho''}_i\right |^2\,\frac{k_i}{2z_{2i}}\right)\right]\exp\left(i{\bm q'}_i{\bm \rho'}_i\right).
\end{eqnarray}
Performing the integrals over the variables ${\bm \rho''}_s$,  ${\bm \rho''}_i$,  ${\bm \rho'}_s$ and ${\bm \rho'}_i$, results in the following expression:
\begin{eqnarray}
\label{eq8}
\left\langle vac\right|\hat{E}^{(+)}({\bm \rho}_s) \hat{E}^{(+)}({\bm \rho}_i) \left|\psi \right\rangle =&&\nonumber\\
{\mathcal A}_2\left({\bm \rho}_i\right)\int d{\bm q'}_s\int d{\bm q'}_iv_{z_0}\left({\bm q'}_s+{\bm q'}_i\right)&&\nonumber\\
\times \exp\left[i\rho^2_s\left(\frac{k_s}{2z_2s}+\frac{k^2_s}{4\alpha_s}\frac{1}{z^2_2}\right)\right]\exp\left[i\rho^2_i\left(\frac{k_s}{2z_2i}+\frac{k^2_i}{4\alpha_i}\frac{1}{z^2_i}\right)\right]
&&\nonumber\\
\times\exp\left[iq^2_s\left(\frac{1}{4\alpha_s}-\frac{z1s}{2k_s}\right)\right]\exp\left[iq^2_i\left(\frac{1}{4\alpha_i}-\frac{z1i}{2k_i}\right)\right]&&\nonumber\\
\times \exp\left(\frac{-ik_s}{2z_{2s}\alpha_s}\right)\exp\left(\frac{-ik_i}{2z_{2i}\alpha_i}\right),
\end{eqnarray}
where the $\alpha_j$, $j=i,\,s$ are given by:
\begin{equation}
\label{eq9}
\alpha_j=\frac{k_j}{2f}-\frac{k_j}{2z_{2j}}.
\end{equation}
Terms in the Eq.(\ref{eq8}) depending only on the variables ${\bm \rho_s}$ and ${\bm \rho_i}$, do not contribute to the integrals over ${\bm q_s}$ and ${\bm q_i}$, and can be omitted, because the coincidence
count rate depends on its square modulus. Assuming that the down-converted fields have the same frequency, and the distances from the crystal to the lenses and from the lenses to the detectors are the same for both modes, we have:
\begin{eqnarray}
\label{eq10}
k_s=k_i=k ; \nonumber\\
\alpha_s=\alpha_i.
\end{eqnarray}
Using the thin lens law $\frac{1}{f}=\frac{1}{I}+\frac{1}{O}$, with $O=z_0+z_1$ and $I=z_2$, we can rewrite Eq.(\ref{eq7}) as:
\begin{eqnarray}
\label{eq11}
\left\langle vac\right|\hat{E}^{(+)}({\bm \rho}_s) \hat{E}^{(+)}({\bm \rho}_i) \left|\psi \right\rangle=&&\nonumber\\
{\mathcal A}_2\left({\bm \rho}_i\right)\int d{\bm q'}_s\int d{\bm q'}_iv_{z_0}\left({\bm q'}_s+{\bm q'}_i\right)&&\nonumber\\
\times \exp \left [i{\mathcal B}\left( q^2_s+q^2_i\right)\right] \exp\left [i\left( q^2_s+q^2_i\right)\right],&&\nonumber\\
\end{eqnarray}
where ${\mathcal B}$ is given by:
\begin{equation}
\label{eq12}
{\mathcal B}=\frac{1}{4\alpha}-\frac{z_1}{k}.
\end{equation}
Introducing the relatives variables
\begin{eqnarray}
\label{eq13}
{\bm u}={\bm q}_s+{\bm q}_i,\nonumber\\
{\bm v}={\bm q}_s-{\bm q}_i,
\end{eqnarray}
we can rewrite Eq.(\ref{eq12}) as:

\begin{eqnarray}
\label{eq14}
\left\langle vac\right|\hat{E}^{(+)}({\bm \rho}_s) \hat{E}^{(+)}({\bm \rho}_i) \left|\psi \right\rangle&=&\nonumber\\
{\mathcal A}_2\left({\bm \rho}_i\right)\int d{\bm u}\int d{\bm v} {\mathcal V}_{z_0}\left({\bm u}\right)\exp \left [i{\mathcal B}\left( \frac{u^2+v^2}{2}\right)\right]&&\nonumber\\
\times \exp\left[-i\frac{O}{I}\left(\frac{{\bm u}+{\bm v}}{2}\cdot{\bm \rho}_s+\frac{{\bm u}-{\bm v}}{2}\cdot{\bm \rho}_i\right)\right]=&&\nonumber\\
\int d{\bm u} {\mathcal V}_{0}\left({\bm u}\right)\exp\left[-i\frac{O}{I}{\bm u}\cdot\left(\frac{{\bm \rho}_i+{\bm \rho}_s}{2}\right)\right ]\nonumber\\
\times \int d{\bm v}\exp\left[i\frac{{\mathcal B}}{2}v^2\right]\exp\left[-i\frac{O}{I}{\bm v} \cdot\left(\frac{{\bm \rho}_i-{\bm \rho}_s}{2}\right)\right].
\end{eqnarray}
The integral over ${\bm u}$ is the Fourier Transform of the angular spectrum of pump beam, while the integral over ${\bm v}$ will add a phase factor, which will be equal to one, when we take the square modulus. Solving these integrals we have the following expression for the coincidence-count rate:
\begin{equation}
\label{eq15}
C\left({\bm \rho}_s,{\bm \rho}_i\right) \propto \left|{\mathcal A}_2({\bm \rho}_i){\mathcal W}_p\left[\frac{-\left({\bm \rho}_s+{\bm \rho}_i\right)}{2}\right]\right|^2.
\end{equation}
 ${\mathcal W}_p= {\mathcal A}_1 {\mathcal E}_0$ is the field distribution of the pump beam immediately after the grating plane, and ${\mathcal E}_0$ is its amplitude before the mask. Assuming that the pump beam is a plane wave, ${\mathcal E}_0=$ const. Thus
\begin{equation}
\label{eq16}
C\left({\bm \rho}_s,{\bm \rho}_i\right) \propto \left|{\mathcal A}_2({\bm \rho}_i){\mathcal A}_1\left[\frac{-\left({\bm \rho}_s+{\bm \rho}_i\right)}{2}\right]\right|^2.
\end{equation}
We see that $C\left({\bm \rho}_s,{\bm \rho}_i\right)$ is given by the product of the transmission functions of the gratings.

\section{Idler-Signal Setup}

\begin{figure}[t]
\centering
\includegraphics*[width=10cm]{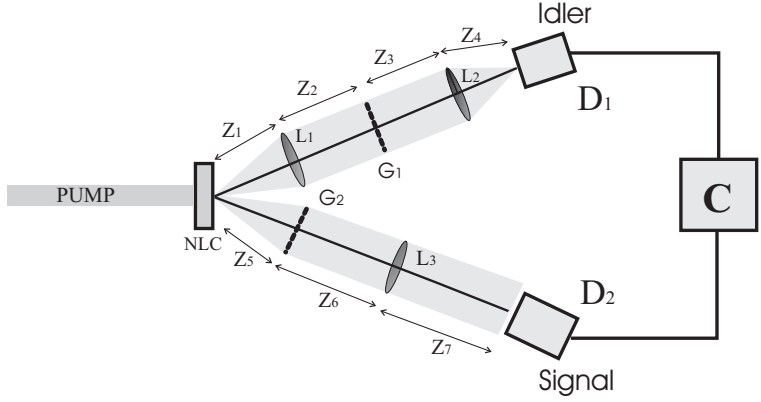}
\vspace{0.25cm}
\caption{Idler-Signal setup.}
\label{f2}
\end{figure}

Now, we will calculate the coincidence count rate for the setup showed in Fig.\ref{f2}. In this case, the two gratings are placed in the signal and idler beams. The resulting moir\'e pattern is produced by the superposition of the correlated images of these objects. We call this configuration idler-signal setup. The field operator for the idler mode,  propagated from crystal to the detection plane, is given by
\begin{eqnarray}
\label{eq17}
\hat{E}^{(+)}_i\left({\bm \rho}_i,Z_i\right)=\int d\, {\bm \rho'}_i\int d\, {\bm \rho''}_i\int d\, {\bm \rho'''}_i
\int d\,{\bm \rho''''}\int d\, {\bm q''''}_i\,\hat{a}\left({\bm q''''}_i\right)\exp\left[ i{\bm q''''}_i \cdot {\bm \rho''''}_i\right]&&\nonumber\\
\times \exp\left[i\left(\left |{\bm \rho}_i-{\bm \rho'}_i\right |^2\,\frac{k_i}{2z_{1i}}\right)\right]{\mathcal T}_1\left({\bm \rho'}_i\right)\exp\left[i\left(\left |{\bm \rho'}_i-{\bm \rho''}_i\right |^2\,\frac{k_i}{2z_{2i}}\right)\right]{\mathcal A}_1\left({\bm \rho''}_i\right)&&\nonumber\\
\times\exp\left[i\left(\left |{\bm \rho''}_i-{\bm \rho'''}_i\right |^2\,\frac{k_i}{2z_{3i}}\right)\right]{\mathcal T}_2\left({\bm \rho'''}_i\right)\exp\left[i\left(\left |{\bm \rho'''}_i-{\bm \rho'''''}_i\right |^2\,\frac{k_i}{2z_{4i}}\right)\right],&&\nonumber\\
\end{eqnarray}
where ${\mathcal T}_1\left({\bm \rho''}_i\right)$ and ${\mathcal T}_2\left({\bm \rho'''}_i\right)$ are the transmission functions of the lenses $L_1$ and $L_2$. ${\mathcal A}_1\left({\bm \rho''}_i\right)$ is the transmission function of the grating $G1$. Solving the integral over ${\bm \rho'}$, ${\bm \rho'''}$ and  ${\bm \rho''''}$, assuming that $z_1=z_4=f$ and  $z_2=z_3=2f$, where $f$ is the focal length of the lenses $L_1$ and $L_2$, it is possible reduce this expression to:
\begin{eqnarray}
\label{eq18}
\hat{E}_i^{(+)}\left({\bm \rho}_i, Z_i\right)=\int\, d{\bm q''''}_i\hat{a}\left({\bm q''''}_i\right)\,\int d {\bm \rho''}_i\,{\mathcal A}_1\left({\bm \rho''}_i\right)\nonumber\\
\times \exp\left[i\frac{q''''^2_i f}{2k_i}\right]\exp\left[-i{\bm \rho''}_i\left(\frac{{\bm \rho}_ik_i}{f}+{\bm q''''}_i\right)\right].
\end{eqnarray}
Again, we have omitted the term that depends only on the variables $\rho_i$.
\par
For the signal mode the field operator is given by:
\begin{eqnarray}
\label{eq19}
\hat{E}^{(+)}_s\left({\bm \rho}_s,Z_s\right)=\int d\, {\bm \rho'}_i\int d\, {\bm \rho''}_s\int d\, {\bm \rho'''}_s
\int d\, {\bm q'''}_i\hat{a}\left({\bm q'''}_i\right)\exp \left[i{\bm q'''}_i\cdot{\bm \rho'''}_i\right]&&  \nonumber\\
\times \exp\left[i\left(\left |{\bm \rho}_s-{\bm \rho'}_s\right |^2\,\frac{k_s}{2z_{5s}}\right)\right]{\mathcal A}_2\left({\bm \rho'}_i\right)\exp\left[i\left(\left |{\bm \rho'}_s-{\bm \rho''}_s\right |^2\,\frac{k_s}{2z_{6s}}\right)\right]{\mathcal T}_3\left({\bm \rho''}_i\right)&&\nonumber\\
\times \exp\left[i\left(\left |{\bm \rho''}_s-{\bm \rho'''}_s\right |^2\,\frac{k_s}{2z_{7s}}\right)\right],&&\nonumber\\
\end{eqnarray}
where ${\mathcal A}_2\left({\bm \rho'}_i\right)$ and ${\mathcal T}_3\left({\bm \rho''}\right)$ are transmission functions of the grating $G_2$ and of the lens $L_4$, respectively. $z_5=f$ and $z_6=z_7=2f$, where f is the focal length of the lens $L_4$. Performing the integrals over the variables ${\bm \rho'}_s$, ${\bm \rho''}_s$ and ${\bm \rho'''}_s$, we have
\begin{equation}
\label{eq20}
\hat{E}_s^{(+)}\left({\bm \rho}_s, Z_s\right)={\mathcal A}_2\left({\bm \rho'}_s\right)\times\int d\, {\bm q'''}_s\hat{a}\left({\bm q'''}_s\right)\exp\left[-i{\bm q'''}_s{\bm \rho}_i\right]\exp\left[\frac{-iq'''^2_s z_1}{zk_s}\right],
\end{equation}
where the terms depending on the variables ${\bm \rho}_s$ were omitted. Using the
expressions for the field operators in Eq.(\ref{eq3}), we have
\begin{eqnarray}
\label{eq21}
\left\langle vac\right|\hat{E}^{(+)}({\bm \rho}_s) \hat{E}^{(+)}({\bm \rho}_i) \left|\psi \right\rangle =&&\nonumber\\
{\mathcal A}_2\left({\bm \rho'}_s\right)\int d\,{\bm \rho''}_i{\mathcal A}_1\left({\bm \rho''}_i\right)\int\,d{\bm q'''}_s\int d{\bm q''''}_i \exp\left[\frac{-iq'''^2_s z_1}{zk_s}\right]\exp\left[-i{\bm q'''}_s \cdot {\bm \rho'}_s\right]&&\nonumber\\
\times \exp\left[-i{\bm \rho''}_i\left(\frac{{\bm \rho}_ik_i}{f}+{\bm q''''}_i\right)\right]{\mathcal V}\left({\bm q'''}_s+{\bm q''''}_i\right),&&\nonumber\\
\end{eqnarray}
where ${\mathcal V}\left({\bm q'''}_s+{\bm q''''}_i\right)$ is the angular spectrum of the pump beam, given in Eq.(\ref{eq2}). Again it is assumed that the signal and idler beams have the same wavelength. Using the change of variables defined in Eq.(\ref{eq13}) we can rewrite Eq.(\ref{eq21})
as
\begin{eqnarray}
\label{eq22}
\left\langle vac\right|\hat{E}^{(+)}({\bm \rho}_s) \hat{E}^{(+)}({\bm \rho}_i) \left|\psi \right\rangle =&&\nonumber\\
{\mathcal A}_2\left({\bm \rho'}_s\right)\int d\,{\bm \rho''}_i{\mathcal A}_1\left({\bm \rho''}_i\right)\,\int\, d{\bm u}\,\int d{\bm v}\, \exp\left[-i{\bm \rho''}_i\frac{{\bm u+v}}{2}\right]&&\nonumber\\
\times \exp\left[i\frac{f}{2k}\left({\bm u\,v}\right)\right]\exp\left[i{\bm \rho'}_s\left(\frac{{\bm u}-{\bm v}}{2}\right)\right]\nonumber\\
\times \exp\left[i\rho''^2_i\frac{k}{2f}\right]\exp\left[-i{\bm \rho''}\frac{\rho_i\,k_i}{f}\right]{\mathcal V}\left({\bm u}\right).&&
\end{eqnarray}
Solving integrals over ${\bm u}$ and ${\bm v}$, we have
\begin{eqnarray}
\label{eq23}
\left\langle vac\right|\hat{E}^{(+)}({\bm \rho}_s) \hat{E}^{(+)}({\bm \rho}_i) \left|\psi \right\rangle=&&\nonumber\\
{\mathcal A}_2\left({\bm -\rho'}_s\right)\times\int d{\bm \rho''}_i\,{\mathcal A}_1\left({\bm \rho''}_i\right){\mathcal V}\left[\frac{k}{f}\left({\bm \rho''}_i-{\bm \rho}_s\right)\right]\exp\left[-i\frac{k{\bm \rho''}_i\,{\bm \rho}_i\,}{f}\right].&&\nonumber\\
\end{eqnarray}
Assuming that the pump beam is a plane wave, we can perform the replacement
\begin{equation}
\label{eq24}
{\mathcal V}\left[\frac{k}{f}\left({\bm \rho''}_i-{\bm \rho}_s\right)\right] \rightarrow \delta\left({\bm \rho''}_i-{\bm \rho}_s\right),
\end{equation}
and we will have the following expression for the coincidence count rate
\begin{equation}
\label{eq25}
C\left({\bm \rho}_s,{\bm \rho}_i\right) \propto \left|{\mathcal A}_2(-{\bm \rho}_s){\mathcal A}_1\left({\bm \rho}_s\right)\right|^2.
\end{equation}
Again, $C\left({\bm \rho}_s,{\bm \rho}_i\right)$ is given by the product of the transmission functions of the gratings.

\section{Discussion and Conclusions}

Let us discuss some aspects of the two configurations investigated.
In the scheme using the transfer of the angular spectrum, called
pump-idler setup, the basic
idea is to prepare a state for the twin beams in which a spatial
structure of a grating is written in the conditional spatial correlations.
The moir\'e is completed by using another grating just before detection.
In this case, there is a clear difference from the ``usual'' (using classical optics) moir\'e. Because of the frequency conversion,
the conditional structure is about two times larger than that placed in the pump
beam. This may represent an improvement over ``usual'' moir\'e, however the
discussion about the improvements of a quantum setup over the classical moir\'e, is complicated by the large number of possible configurations
for a moir\'e. A complete discussion is beyond the scope of this work.

From Eq.(\ref{eq16}), it is seen the conditional character of the coincidence
spatial distribution, through the dependence on the sum of the coordinates
of the signal and idler detectors.

The scheme using one grating in each one of the beams, called signal-idler,
is easily understood
in terms of the Klyshko's advanced wave picture\cite{advanwaves}. Idler detector works as it was the
light source, shining light onto the idler grating. The image of the grating
is projected onto the signal grating, as if the light ´´emitted" from the idler detector was reflected into the signal path by the non-linear crystal. Finally, a lens is
used to image the plane of the signal grating onto the signal detection plane,
resulting in the moir\'e. According to Eq.(\ref{eq25}), the coincidence count
rate depends only on the signal detector coordinates. This may seem surprising
at first sight, but it can be understood from the advanced wave interpretation,
because the idler detector, which is playing the role of the source, is placed
at the focal plane of a lens. Therefore, this detector is not sensitive to
position displacements of the idler detector. In Eq.(\ref{eq25}), the detector has been
considered a point detector, and because of the spatial Fourier transform realized
by the lens, it is working as a filter for the spatial frequencies. In this
scheme, possible improvements over classical moir\'e is less evident,
however this does not mean that improvements are impossible.

In both pump-idler and signal-idler schemes, there is one important difference
from any classical implementation. Because of the non-local correlations
between signal and idler photons, the information concerning the moir\'e depends
on both signal and idler detections. Therefore, one can delay the detection
of one of the photons. In the meantime, the non detected photons will be in
a superposition state, while in a classical scheme, this would not be possible.

In conclusion, we have presented the calculation of the coincidence count rate
for two implementations of moir\'e fringes using quantum images. The theoretical results also
bring to attention special features of each scheme, such as the conditional
character of the scheme pump-idler and the dependence only on the idler coordinates
for the signal-idler scheme. We believe our study will assist for practical
use of the quantum moir\'e fringes.

\begin{acknowledgments}
We acknowledge Dr. S. P. Walborn for reading this manuscript, and the financial support provided by Brazilian agencies CNPq, PRONEX, CAPES, FAPERJ, FUJB and the Milenium Institute for Quantum Information.
\end{acknowledgments}




\end{document}